\documentclass{article}

\usepackage[pdftex]{hyperref}
\usepackage{subfigure}
\usepackage{graphicx}
\usepackage{amsmath}
\usepackage{natbib}
\usepackage{helvet,soul}
\usepackage[font=small]{caption}
\usepackage{xcolor}
\usepackage[english,spanish]{babel}
\usepackage{authblk}

\begin{document}

\title{Study of young stellar objects around SNR G18.8+0.3}

\author[1]{M. Celis Pe\~na}
\author[1]{S. Paron}
\affil[1]{Instituto de Astronom\'ia y F\'isica del Espacio (IAFE)}

\maketitle

\selectlanguage{english} 

\begin{abstract}

In recent works, through observations of molecular lines, we found that the supernova remnant (SNR) G18.8+0.3 is interacting with a molecular cloud towards its southern edge. Also it has been proven the presence of several neighboring HII regions very likely located at the same distance as the remnant. The presence of dense molecular gas and the existence of shock fronts generated by the SNR and HII regions make this region an interesting scenario to study the population of young stellar objects. Thus, using the most modern colour criteria applied to the emission in the mid-infrared bands obtained from IRAC and MIPS on board Spitzer, we characterized all the point sources lying in this region. We analyzed the spectral energy distributions of sources that show signs of being young stellar objects in order to confirm their nature and derive stellar parameters. Additionally, we present a map of the $^{12}$CO J=3-2 emission obtained with the ASTE telescope towards one of the HII regions embedded in the molecular cloud. The molecular gas was studied with the aim to analyze whether the cloud is a potential site of star formation.
\end{abstract}

\selectlanguage{spanish} 
\begin{abstract}

Recientemente, a trav\'es de observaciones de l\'ineas moleculares hemos comprobado que el remanente de supernova (RSN) G18.8+0.3 se encuentra interactuando con una nube molecular hacia su borde sur. Tambi\'en se ha comprobado la presencia de varias regiones HII vecinas asociadas al gas molecular, las cuales muy probablemente se encuentran a la misma distancia que el remanente. La presencia de gas molecular denso, y la existencia de frentes de choque, generados tanto por el RSN  como por las regiones HII, convierten a esta regi\'on en un interesante escenario en donde estudiar la poblaci\'on de objetos estelares j\'ovenes. De esta manera, utilizando los criterios de color m\'as modernos aplicados a la emisi\'on de las bandas del infrarrojo medio obtenidas de los instrumentos IRAC y MIPS a bordo del sat\'elite Spitzer, se caracterizaron todas las fuentes puntuales que yacen en esta regi\'on. A aquellas  que presentan indicios de ser objetos estelares j\'ovenes se les analiz\'o su distribuci\'on espectral de energ\'ia, con el fin de confirmar su naturaleza y extraer varios par\'ametros estelares. De manera adicional se presenta un mapa de la emisi\'on del $^{12}$CO J=3-2 obtenido con el telescopio ASTE hacia una de las regiones HII vecinas al RSN. Se realiz\'o una caracterizaci\'on del gas molecular con el fin de analizar si la nube es un potencial sitio de formaci\'on estelar activa.
\end{abstract}

\selectlanguage{english}


\section{Introduction}
\label{S_intro}

\noindent 

\begin{figure}[!ht]
  \centering
  \includegraphics[width=0.45\textwidth]{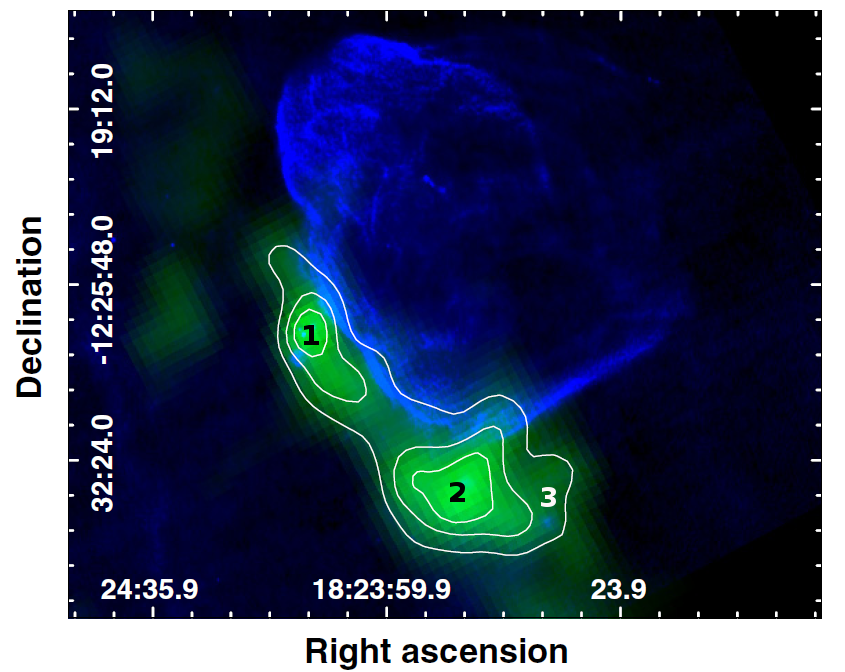} 
  \caption{\label{figintro} SNR G18.8+0.3 in radio continuum emission at 20 cm (blue) and the integrated $^{13}$CO J=1--0 emission extracted from the Galactic Ring Survey (green).}
\end{figure}

The interaction between the supernova remnant (SNR) G18.8+0.3 and a molecular cloud was analyzed in recent works. The molecular gas belonging to regions 1 and 2 (see Figure \ref{figintro}) was studied in \citet{paron2012} and \citet{paron2015}, respectively. The presence of abundant molecular gas, several embedded HII regions and the SNR, makes this region an interesting scenario to study possible triggered star formation. Thus, in this work we present a study of the distribution of the young stellar objects (YSOs) in the SNR surroundings. In addition, we study the molecular gas towards the named region 3 in Figure \ref{figintro}, which is related to the HII region G018.584+00.344.

\section{Results}
\subsection{Young stellar objects}

\begin{table}
\centering
\caption{Stellar parameters obtained from the SEDs.}
\tiny
\begin{tabular}{cccccccccccc}
\hline\hline\noalign{\smallskip}
\!\!Source & \!\!\!\!$M_\star$ & \!\!\!\!Range & \!\!\!\!Age & \!\!\!\!Range & \!\!\!\!${\dot M}_\text{env}$ &\!\!\!\!Range &\!\!\!\!$L$ &\!\!\!\!Range & \!\!\!\!$\chi^2_\text{best}$ &\!\!\!\!Satisfying &\!\!\!\!Stage\!\!\!\!\\

& \!\!\!\![$M_\odot$] & \!\!\!\!$M_\star$ & \!\!\!\![$10^5$yr] & \!\!\!\!Age & \!\!\!\![$10^{-3}M_\odot \text{yr}^{-1}$] & \!\!\!\! ${\dot M}_\text{env}$ & \!\!\!\![$10^3 L_\odot$] &\!\!\!\!$L$ &\!\!\!\! & \!\!\!\! models & \\
\hline\noalign{\smallskip}
1  & 16.3   & --   			 & 12.5   & --		    & 0      & --        & 26.7   & --		    & 52.4   & 10	 	& III 	\\
2  & 10.0   & 5.7--15.2	     & 15.5   & 11.4--47.0  & 0      & --        & 6.2    & 1.2--21.6   & 1.0    & 305   	& II	\\
3  & 9.1    & 5.9--13.6	 	 & 0.01   & 0.01--37.7  & 1.4    & 0--6.6    & 0.2    & 0.9--11.8   & 3.2    & 23   	& 0-I	\\
4  & 7.8    & 1.0--11.5		 & 31.4   & 0.01--67.6  & 0      & 0--0.5    & 2.5    & 0.2--9.8    & 0.7    & 706   	& II	\\
5  & 18.8   & 17.8--19.9	 & 12.8   & 12.8--14.5  & 0      & --        & 39.6   & 34.1--46.2  & 63.2   & 15		& III	\\
6  & 18.5   & 16.4--20.9	 & 13.1   & 13.0--14.3  & 0      & --        & 37.9   & 26.9--50.1  & 69.3   & 27		& III	\\
7  & 8.0    & --		 	 & 0.4    & --    		& 0.5    & --  		 & 0.9    & --    		& 15.1   & 2		& 0-I	\\
8  & 13.9   & 6.7--16.3 	 & 15.5   & 0.02--24.6  & 0      & 0--0.6    & 17.1   & 0.9--26.7   & 3.7    & 203		& II	\\
9  & 4.3    & 3.2--6.9	 	 & 20.5   & 0.8--78.1   & 0      & 0--0.9    & 0.2    & 0.1--1.6    & 0.1    & 461		& II	\\
10 & 7.0    & 7.0--9.8	 	 & 18.5   & 10.9--43.8  & 0      & --        & 2.0    & 1.9--5.9    & 10.2   & 98		& II	\\
11 & 14.4   & 8.7--23.8	 	 & 0.06   & 0.01--24.6  & 1.2    & 0--4.4    & 7.6    & 4.0--30.7   & 1.3    & 71		& 0-I	\\
12 & 8.1    & 7.2--10.8	 	 & 21.7   & 0.6--32.1   & 0      & 0--1.8    & 3.0    & 2.1--8.0    & 12.4   & 32		& II	\\
13 & 16.3   & --	 		 & 12.5   & -- 		    & 0      & --        & 26.7   & -- 		    & 77.4   & 10		& III	\\
14 & 18.5   & 18.5--20.6	 & 13.1   & 13.0--13.1  & 0      & --        & 37.9   & 37.9--50.1  & 26.0   & 16		& III	\\
15 & 15.6   & 9.9--19.9	 	 & 16.3   & 11.8--32.6  & 0      & --        & 23.4   & 6.2--45.9   & 47.1   & 33		& III	\\
16 & 16.3   & --	 		 & 12.5   & -- 		    & 0      & --        & 26.7   & -- 		    & 148.7  & 9		& II	\\
17 & 13.9   & 13.9--19.9	 & 18.2   & 13.8--18.2  & 0      & --        & 19.0   & 19.0--46.2  & 39.7   & 14		& II	\\
18 & 15.6   & 13.6--19.9	 & 16.3   & 11.8--16.3  & 0      & --        & 23.4   & 15.8--45.9  & 39.2   & 19		& III	\\
19 & 16.9   & 15.7--18.8	 & 12.4   & 12.4--19.3  & 0      & --        & 29.3   & 23.7--39.6  & 74.6   & 34		& II	\\
20 & 4.8    & 4.4--7.2	     & 9.5    & 0.6--60.3   & $10^{-4}$ & 0--0.6 & 0.4    & 0.1--1.6    & 0.0    & 340		& II	\\
21 & 3.9    & 3.1--5.7	  	 & 52.1   & 3.6--92.6   & 0      & 0--0.05   & 0.1    & 0.1--0.7    & 0.7    & 205		& II	\\
22 & 11.5   & 7.2--13.7	 	 & 19.5   & 11.6--38.4  & 0      & --        & 9.8    & 2.5--16.3   & 4.3    & 58		& II	\\
23 & 18.5   & --			 & 13.1   & -- 		    & 0      & --        & 37.9   & --		    & 34.3   & 10		& III	\\
24 & 9.6    & 9.4--15.8	 	 & 30.3   & 18.7--30.3  & 0      & --        & 5.5    & 5.1--24.4   & 36.2   & 25		& III	\\
25 & 6.2    & 4.6--7.9	 	 & 31.1   & 10.9--59.8  & 0      & --        & 1.0    & 0.4--2.7    & 1.73   & 219		& II	\\
26 & 9.7    & 7.4--10.8	 	 & 17.8   & 13.6--32.6  & 0      & --        & 5.7    & 2.1--8.1    & 83.3   & 52		& III	\\
27 & 16.9   & 12.0--17.9	 & 12.4   & 12.4--22.2  & 0      & --        & 29.3   & 11.5--34.4  & 53.6   & 32		& II	\\
28 & 16.4   & 16.3--18.6	 & 14.3   & 12.5--14.3  & 0      & --        & 26.9   & 26.7--38.3  & 60.6   & 28		& III	\\
29 & 12.0   & 11.9--12.0	 & 22.2   & 15.3--22.2  & 0      & --        & 11.5   & 11.3--12.1  & 74.2   & 6		& II	\\
30 & 9.9    & 9.9--15.6	 	 & 32.6   & 13.8--32.6  & 0      & --        & 6.2    & 6.2--23.4   & 42.7   & 27		& III	\\   
\hline
\multicolumn{12}{l}{\tiny A "--" in the range columns means that all the satisfaying models share the same value as the best-fit model.} 
\end{tabular}
\label{tablaSED}
\end{table}

\begin{figure}[!ht]
  \centering
  \includegraphics[width=.45\textwidth]{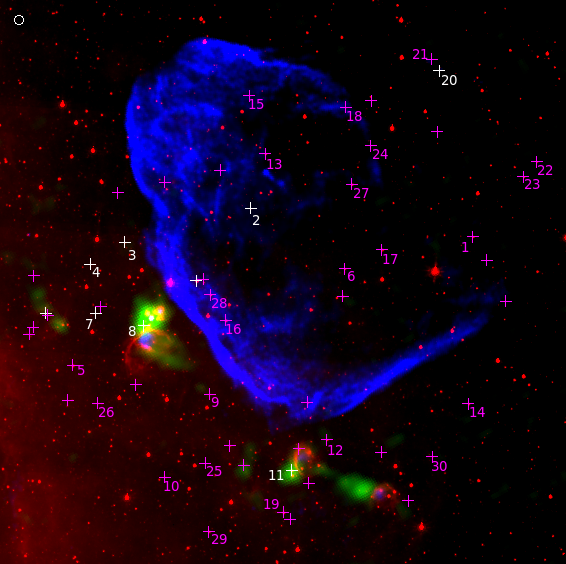}
  \caption{\label{estrellas_jovenes} SNR G18.8+0.3 in radio continuum emission at 20 cm (blue), submillimeter continuum at 870 $\mu$m from ATLASGAL (green), and 8 $\mu$m emission from IRAC-Spitzer (red). White and magenta crosses are Class I and II sources according \citet{gutermuth2009}.}
\end{figure}

We search for YSO candidates towards a region of about 5$^{\prime}$ in size around the SNR G18.8+0.3. Using the GLIMPSE Point-Source Catalog acquired by Spitzer-IRAC (at 3.6, 4.5, 5.8 and 8 $\mu$m) we selected sources following the criteria presented in \citet{gutermuth2009}. 
We found 9 Class I objects and 47 Class II objects, white and magenta crosses in Figure \ref{estrellas_jovenes}, respectively, which are displayed over a three-colour image with the radio continuum emission at 20 cm presented in blue, the submillimeter continuum emission at 870 $\mu$m in green, and the 8 $\mu$m emission in red. It can be appreciated that some YSO candidates are related to cold dust condensations mapped by the submillimeter continuum. Thirty of these objects have their fluxes measured at 24 $\mu$m with MIPS-Spitzer, and thus for these sources we performed an analysis of its spectral energy distribution (SED) using the tool developed by \citet{robitaille2007} and available online \footnote[1]{http://caravan.astro.wisc.edu/protostars/}

To build the SEDs models we used the Spitzer-IRAC fluxes measured in the bands 3.6, 4.5, 5.8, 8.0 $\mu$m and the flux at 24 $\mu$m measured with Spitzer-MIPS. Using the criteria presented in \citet{robitaille2006}, the YSO candidates can be classified in:

\begin{itemize}

\item[•] Stage 0-I: objects with significant infalling envelopes and possibly disks. This stage is defined by $\dot{M}_{env}/M_\star > 10^{-6} yr^{-1}$, where $\dot{M}_{env}$ is the envelope accretion rate, and $M_\star$ is the central source mass.
\item[•] Stage II: objects with optically thick disks (and possible remains of a tenuous infalling envelope). This stage is defined by $\dot{M}_{env}/M_\star < 10^{-6} yr^{-1}$ and $M_{disk}/M_\star > 10^{-6} yr^{-1}$, where $M_{disk}$ is the disk mass. 
\item[•] Stage III: objects with optically thin disks. This stage is defined by  $\dot{M}_{env}/M_\star < 10^{-6} yr^{-1}$ and $M_{disk}/M_\star < 10^{-6} yr^{-1}$.
\end{itemize}

Table \ref{tablaSED} presents some stellar parameters obtained from the SED of each source. The parameters from the best fit model and their ranges obtained from all the satisfying models are presented. To do this, we selected the models that satisfied the condition $\chi^2 - \chi^2_\text{best} < 3N$, where $\chi^2_\text{best}$ is the minimum value of the $\chi^2$ among all models, and $N$ is the number of input data fluxes. Additionally the source stage is included in the last column of the table.

Taking into account our photometric and SED analysis of the sources and their spacial distribution towards the SNR and the associated molecular cloud, we conclude that there is no evidence that allow us to infer triggered star formation in the region. Additionally, performing a SED to the embedded HII regions (see \citet{paron2012,paron2015}), it is concluded that they are located at the same distance as the SNR (about 8 kpc). With this in mind, and assuming that many of the sources analyzed here are at the same distance, we conclude that we are studying a region populated by many massive stars in different evolutive stages.

\subsection{Molecular gas analysis towards the HII region G018.584+00.344}

\begin{figure}[!ht]
  \centering
  \includegraphics[width=.45\textwidth]{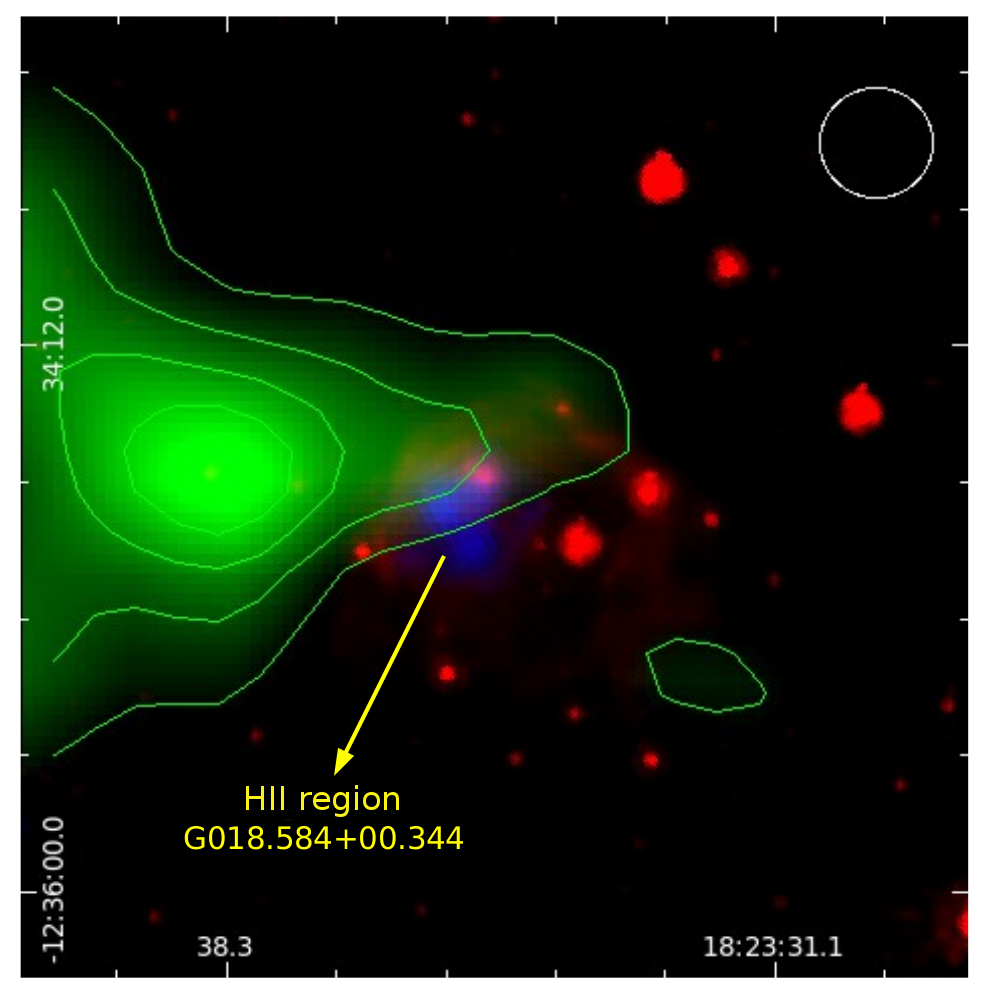}
  \caption{\label{region3} Radio continuum emission at 20 cm (blue), IR emission at 8 $\mu$m (red), and the $^{12}$CO J=3--2 integrated between 15 and 30 km s$^{-1}$ emission (green), with contours levels of 30, 40, 50 and 60 K km s$^{-1}$. The beam of the molecular observations is presented at the top right corner.}
\end{figure}

Using molecular data obtained with the Atacama Submillimeter Telescope Experiment (ASTE) (see \citet{paron2015} for a description of the observations), we investigated the interstellar medium around the HII region G018.584+00.344 which lies southwards the SNR G18.8+0.3 (region 3 in Fig.\ref{figintro}). Figure \ref{region3} is a three-colour image where the radio  continuum emission at 20 cm is presented in blue, the 8 $\mu$m emission in red, and the integrated $^{12}$CO J=3--2 emission obtained with ASTE in green. The molecular peak, located estwards the HII region G018.584+00.344, coincides with an ATLASGAL submillimeter continuum source, showing an excellent correlation between the molecular gas and cold dust.

From the molecular feature defined by the 50 K km s$^{-1}$ contour in Fig.\ref{region3} we estimated the molecular mass from three independent ways: using the measured CO luminosity, from the cold dust emission, and through tha virial theorem. The results are:

\begin{align*}
	 M_\text{COlum} &\sim 8.5 \times 10^3M_\odot\\
	 M_\text{dust}  &\sim 5.7 \times 10^3M_\odot\\
	 M_\text{vir}   &\sim 8.2 \times 10^3M_\odot
\end{align*}

From the comparison between the obtained mass values, it is concluded that the molecular clump is gravitationally bound, and taking into account that there is not any YSO candidate embedded in it, we suggest that it is a quiescent molecular clump.

\section{Acknowledgement}
S.P is member of the \textit{Carrera del investigador cient\'ifico} of CONICET, Argentina. M.C.P is a doctoral fellow of CONICET, Argentina. This work was partially supported by Argentina grants awarded by UBA (UBACyT), CONICET and ANPCYT. 
%


\bibliographystyle{plainnat}
\small
\bibliography{referencias}
 
\end{document}